\documentclass[pre,floatfix,twocolumn, 10pt]{revtex4-1} 
\pdfoutput=1
\usepackage{graphicx}
\usepackage{amssymb}
\usepackage{natbib}
\usepackage{dcolumn}
\usepackage{bm}
\usepackage{color}
\usepackage[colorlinks=true, linkcolor=blue, citecolor=blue]{hyperref}
\usepackage{subfigure}
\usepackage{graphicx,psfrag,xspace}
\usepackage{color}
\usepackage{amssymb,amsfonts,amsmath}
\usepackage{setspace}

\definecolor{red}{rgb}{1,0,0}
\definecolor{blue}{rgb}{0,0,1}
\definecolor{green}{rgb}{0.5,0.7,0.5}

\renewcommand{\d}{{\rm d}}

\newcommand{\lp}{\left(}
\newcommand{\rp}{\right)}
\newcommand{\req}[1]{(Eq.~\ref{Eq:#1})}

\begin{document}

\author{Fran\c cois P. Landes}
\affiliation{The Abdus Salam International Center for Theoretical Physics, Strada Costiera 11, 34014 Trieste, Italy\\ Laboratoire de Physique Th\' eorique et Mod\` eles Statistiques (UMR CNRS 8626), Universit\' e Paris-Sud, Orsay, France}

\author{Alberto Rosso}
\affiliation{Laboratoire de Physique Th\' eorique et Mod\` eles Statistiques (UMR CNRS 8626), Universit\' e Paris-Sud, Orsay, France}

\author{E. A. Jagla} 
\affiliation{Centro At\'omico Bariloche and Instituto Balseiro (UNCu), 
Comisi\'on Nacional de Energ\'{\i}a At\'omica, (8400) Bariloche, Argentina}

\title{Frictional dynamics of viscoelastic solids driven on a rough surface}

\begin{abstract} 
We study the effect of viscoelastic dynamics on the frictional properties of a (mean field) spring-block system pulled on a rough surface by an external drive.  When the drive moves at constant velocity $V$, two dynamical regimes are observed: at fast driving, above a critical threshold $V^c$, the system slides at the drive velocity  and  displays a  friction force with {\em velocity weakening}. Below $V^c$ the steady sliding becomes unstable and a stick-slip regime sets in. In the  slide-hold-slide driving protocol, a peak of the friction force appears after the hold time and its amplitude increases with the hold duration. These observations are consistent with the frictional force encoded phenomenologically in the rate-and-state equations. Our model gives a microscopical basis for such macroscopic description.
\end{abstract}

\maketitle

\section{Introduction}

Friction is behind many phenomena of our every day life experience such as
the adhesion of car tires on the road \cite{Persson2005a, Persson2011},
 the sound emitted by a violin  \cite{Smith2000, Serafin1999},
or the wear of human articular joints \cite{Lee2013}.
Often, in these situations, the two bodies in contact
 may display discontinuous dynamics, the so-called stick-slip dynamics \cite{Baumberger1995, Baumberger2006}, in which periods of rapid movement (slip) are followed by periods of relative rest (stick). 
At the macroscopic scale the stick-slip motion is justified by the empirical difference observed between the static and dynamic friction coefficient, but 
a general explanation of this phenomenon from first principles is not yet available. 
The introduction of visco-elastic effects in the dynamics can be the key for a microscopic theory of friction.
In fact, we have recently shown that viscoelastic solids sliding on a rough substrate can display a stick-slip instability \cite{Jagla2014a}. Interestingly it was also pointed out that visco-elasticity is at the origin of the observed precursors, the micro-slips occurring at the onset of slip \cite{Radiguet2015, DiBartolomeo2012, Amundsen2011}.

\begin{figure}
\includegraphics{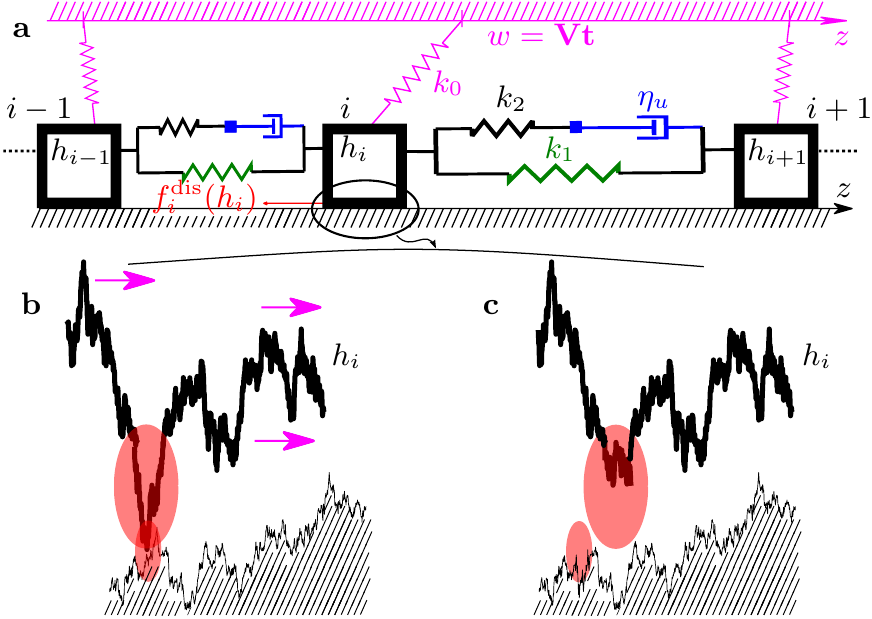}
\caption{
(Color online)
\textit{Sketch of the viscoelastic interface model.}
(\textbf{a}): 
the interface consists in blocks
(labelled $i,i+1, \dots$) 
located at the positions $h_i,h_{i+1},\dots$ (empty squares) and bound together via a combination of springs ($k_1, k_2$) and dashpots ($\eta_u$).
Driving is performed via springs $k_0$ linked to the position $w=Vt$. 
(\textbf{b},\textbf{c}): the asperities that provide the contact (highlighted with ellipses) hinder the sliding of the block on its substrate, with a random force $f^\text{dis}_i(h_i)$ acting as a microscopic static friction force.
(\textbf{c}): the solid sled on the substrate and the asperities under stress are changed.
\label{Fig:model}
}
\end{figure}

A scheme of the model studied in \cite{Jagla2014a} is reproduced for convenience  
in Fig.~\ref{Fig:model}. It consists of an ensemble of interacting blocks driven at velocity $V$ on a rough substrate.
The macroscopic friction arises from the real area of solid-substrate contact, which consists in the junctions between asperities.
When the block is stuck on the rough substrate, the elastic energy of the spring $k_0$ slowly accumulates over time, and is released when junctions break, letting the block move. 
The rupture of a single junction can trigger further rupture, with a characteristic time $\tau_0$ that characterizes the dynamics of the block-substrate contact.
In the quasi-static limit ($V\to 0^+$), when the driving time scale $\tau_D$ is very slow compared to $\tau_0$, the chain of events triggered by a single rupture can be very large and is called an \textit{avalanche}.

 In the case of purely elastic interactions between the blocks, the macroscopic friction force, namely the average elongation of the spring $k_0$, is the control parameter of the depinning transition, a second order transition between a pinned phase and a phase where the system slides at the driving velocity $V$ \cite{Kardar1998,Fisher1998, Sethna2001, Zapperi1998a,Rosso2002}.
Visco-elastic interactions \cite{Schwarz2003} change the nature of the moving phase, inducing hysteretic behavior, as it was shown by Marchetti {\em et al.}\cite{Marchetti2000, Marchetti2005} in the context of the plastic depinning transition of a vortex lattice. 

In the quasi-static driving limit ($\tau_0 \ll\tau_D$), the presence of visco-elasticity (with a characteristic time $\tau_u$) has been shown to produce a rich phenomenology of the avalanche dynamics, with main shocks, aftershocks and spatio-temporal correlations similar to that observed in seismology \cite{Jagla2010a,Jagla2010,Bottiglieri2010, Rosso2012, Papanikolaou2012,Jagla2014a}.
When the visco-elastic relaxation is very fast compared to the driving ($\tau_u \ll \tau_D$), these system-size avalanches can be understood as a stick-slip dynamics of the mean field model \cite{Jagla2014a}.

In the present paper we study the case where the driving rate $\tau_D$ is of the same order of magnitude as the viscoelastic time scale $\tau_u$, but is still very slow compared to the duration of each single avalanche: $\tau_0 \ll \tau_u \simeq \tau_D$. 
We show that in this limit, the system displays the basic features of dry friction.
In particular when a uniform driving is applied, we observe a transition between a stick-slip regime (slow driving, $\tau_u \ll \tau_D$) and a steady-sliding regime (fast driving, $\tau_u \gg \tau_D$).
When the driving velocity is not constant, as in the case of the slide-hold-slide protocol, our results display qualitatively the kind of behavior observed in the experiments \cite{Kilgore1993}, most remarkably, the existence of a peak in the friction force after the hold period, which becomes larger as the hold time increases. 
It must be noted that 
this kind of experiments on friction dynamics are usually modelled using the so-called rate-and-state equations \cite{Dieterich1979, Ruina1983}, which incorporate the fact that the friction force does not simply depend on the relative velocity between the two nominal surfaces but also depends on the history of the contact, via a macroscopic ``state'' variable $\theta$. 
In our model this dependence is encoded in the state of the microscopic visco-elastic elements. In this way our model gives a microscopical basis for the rate-and-state description of friction.

The paper is organized as follows.
 In section \ref{sec:model} we briefly present the model.
 In section \ref{sec:constant_drive_proto} we study the case of uniform driving, and complete this study with the slide-hold-slide protocol in section \ref{sec:slide_hold_slide}.
 We then interpret our results in terms of rate-and-state formalism in section \ref{sec:RSF_analysis} and conclude in section \ref{sec:conclu}.

\section{Model}
\label{sec:model}

\begin{figure}
\includegraphics[width=0.48 \textwidth]{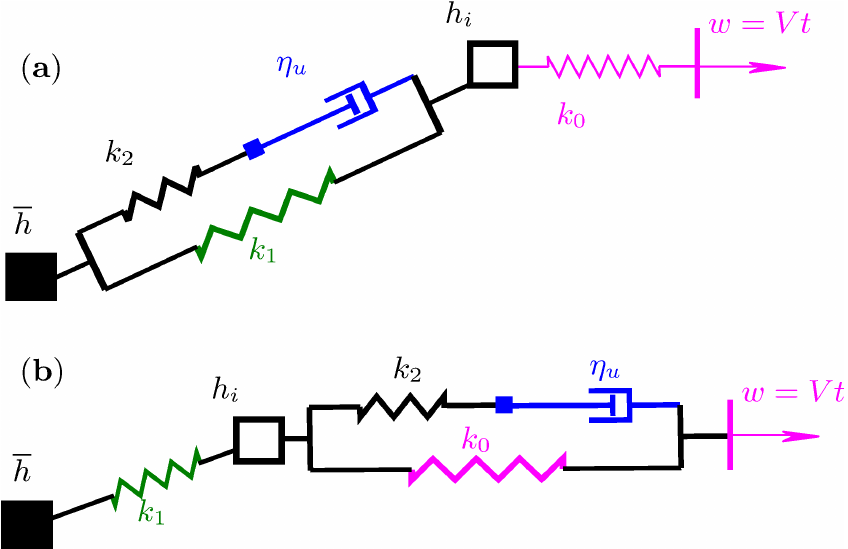}
\caption{
(Color online)
(a): Mean field model of Eqs.~(\ref{Eq:fast1}) and
(\ref{Eq:relaxation1}).  (b): Equivalent model if the solution is
stationary.
}
\label{A3_1}
\end{figure}

The equations of motion for the model pictured in Fig.~\ref{Fig:model} are derived  in app.~\ref{sec:pedestrian}:
\begin{align}
\eta_0 \partial_t h_i & =   k_0(w-h_i) + 
k_1 \nabla^2 h_i +  f_i^\text{dis}(h_i)  \notag  \\
& + k_2 \nabla^2 h_i  - k_2 u_i  \label{Eq:fast1} \\
\eta_u \partial_t u_i & = k_2 \nabla^2 h_i  -k_2 u_i, \label{Eq:relaxation1}
\end{align}
Here $h_i$ is the position  of the block $i$, $w=Vt$ is the driving term in the uniform driving protocol 
and $\eta_0$ is the damping coefficient for the sliding of blocks relatively to the substrate.
The internal variable $u_i$ accounts for the dashpots elongation, which induce the damping 
of the elastic force,  $k_2 (\nabla^2 h_i -u_i)$,  acting on the block $i$.
The characteristic response time of blocks is $\tau_0 = \eta_0/ \max(k_0,k_1,k_2)$, while the characteristic re-adjustment time for the dashpots is $\tau_u=\eta_u/k_2$.

 The random force  $ f_i^\text{dis}(h_i)$  mimics the contacts between the asperities of the block and those of the surface, and acts as a microscopic static friction force.
When a contact is broken, two things happen: (i) the block moves and redistributes stress to neighbouring blocks, (ii) the asperities involved in the junction are renewed, as sketched in Fig.~\ref{Fig:model}\textbf{c}.
This renewal process makes the asperity landscapes different for each block.
For simplicity, we consider that the random force $f_i^\text{dis}(h_i)$ acting on block $i$ is completely independent from that acting on other blocks.
   
Eqs.~(\ref{Eq:fast1}) and (\ref{Eq:relaxation1}) were introduced in \cite{Jagla2014a} and studied  in $d=2$ and in mean field for $V=0^+$. 
Here we extend the mean field approximation (which corresponds to replace the laplacian $\nabla^2 h_i $ with  $ \overline{h}-h_i$, where $\overline{h}$ is the average location of the blocks, see Fig. \ref{A3_1}(a)) to the finite velocity case.
Note that this model couples the $h_i$'s with $\overline h\equiv \sum_i h_i/N$, i.e. it actually represents $N$ times the $k_2, \eta_u, k_1$ units.
As it is usual in the mean field case, the values of the parameters are thus scaled as $k_2/N, \eta_u/N, k_1/N$, to ensure the extensive character of the total energy.
We compute the macroscopic friction force
\begin{align}
\sigma \equiv k_0(w-\overline{h})
\end{align}
accumulated in the system for different driving protocols and adopt the so-called ``narrow wells'' approximation \cite{Fisher1998}. 
In this scheme, the disorder $f^{\text{dis}}_i(h_i)$ is modelled as a collection of narrow pinning wells representing impurities, with spacings $z$ distributed as $g(z)$ and with average $\overline{z}= \int_0^\infty z g(z) \d z$. 
As the wells are very narrow, the disorder force $f^{\text{th}}_i(h_i)$ that derives from this potential is $0$ everywhere except for countably many points. Within this approximation each block is pinned in a single narrow well (see the figure in app.~\ref{sec:pedestrian}). 
For simplicity we consider narrow wells of constant depth (and shape), namely, the random thresholds are constant, $f^ \text{th}_i =\text{const.}=1$. 

Our model provides a description of the interface between two solid surfaces where all dynamical properties are concentrated in one of them, and the other (the substrate) is taken as inert. It has to be emphasized that the blocks have to be considered as microscopic single contacts between the surfaces, and as such they are individually given a rather trivial and time-independent interaction law with the substrate. In addition, the $k_1$, $k_2$, and $\eta_u$ elements (and to a certain extent also the $k_0$ springs) must be considered as part of the surface itself, the whole picture in Fig. \ref{A3_1} thus representing a small part of the two solid surfaces in contact.

Note that here and in the rest of the paper, we give all quantities in dimensionless form.
In order to restore physical units we need to re-introduce the units of the fundamental quantities: distance along the $h$ direction ($\bf z$), along the surface ($\bf x$), force ($\bf f$), and time ($\bf t$), so for instance the previous unit value of $f^{th}$ means $f^{th}=1{\bf f}$, spring constants $k_1, k_2$ are given in units of ${\bf f}{\bf x}^2/{\bf z}$, $k_0$ is in units of  ${\bf f}/{\bf z}$, etc.

\begin{figure}
\includegraphics[width=0.48 \textwidth]{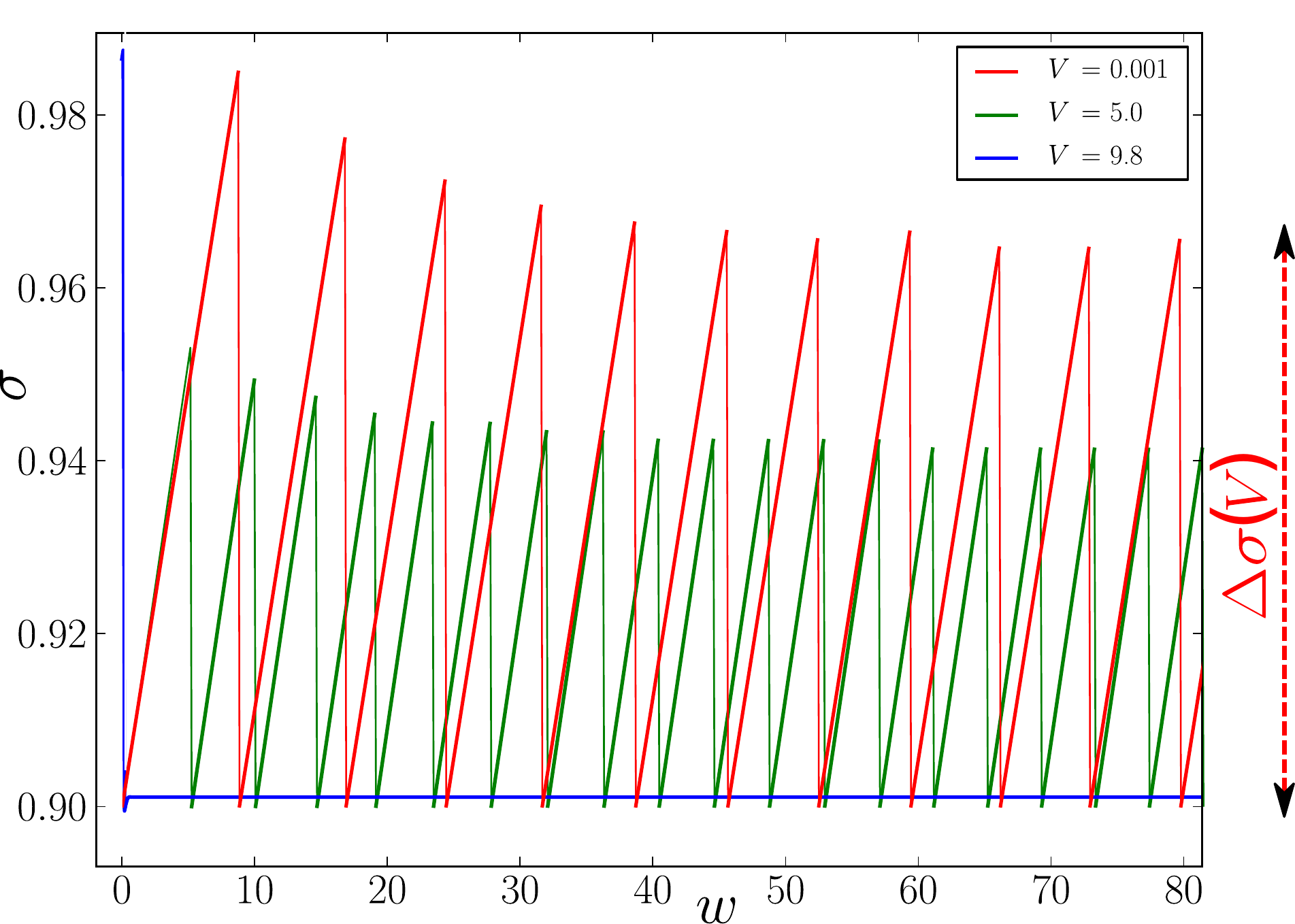}
\caption{
(Color online)
\textit{Evolution of the stress $\sigma$ over time for three velocities. }
At slow driving $V=0.001$ (large amplitude), $V=5$ (small amplitude) the stress oscillates periodically, with an amplitude that we denote $\Delta \sigma (V)$.
At faster driving $V=9.8$ (lower curve, almost always constant), the stress reaches a stationary value after a very short transient.
The precise choice of initial condition only impacts the transient regimes.
We used $k_0=0.01$, $k_1=0.1$, $k_2=0.9$ and $\overline{z}=0.1$.
\label{Fig:SigmaVSt}
}
\end{figure}

\section{Uniform driving protocol}
\label{sec:constant_drive_proto}

At long times, under a driving performed at constant velocity $V$, the solution of Eqs.~(\ref{Eq:fast1}) and (\ref{Eq:relaxation1})  becomes independent of the initial condition and, in mean field, 
two phases are observed: for small $V$ we find a limit cycle solution that corresponds to a stick-slip phase, whereas for large $V$ a stationary solution exists. 
The stationary value of the friction force, $\sigma(V)$, can be computed analytically for $g(z)=\delta(z-\overline{z})$ using the crucial remark that if a steady sliding regime exists then in this regime $\overline{h}$ moves with velocity $V$. In particular, $\overline{h}=V t - \sigma(V)/k_0$.
In the stationary regime the $k_2$-plus-dashpot branch shown in Fig.~\ref{A3_1}(a)
can also be thought of as connecting $h_i$ with $w =V t$, instead of  $\overline{h}$, as indicated in Fig. \ref{A3_1}(b).
 This is so because the additional stretching induced by the time independent shift, $ w-\overline{h}=  \sigma(V)/k_0$, is quickly absorbed by the dashpot, without altering the forces in any manner.

 The model in Fig. \ref{A3_1}(b) coincides with the one studied by Dobrinevski, Le Doussal, and Wiese in   \cite{Dobrinevski2013} and can be solved exactly. In particular the friction force writes (see appendix \ref{stationary}) 
\begin{equation}
\sigma(V)=f^{th}-\left[V\eta_u-\frac{\overline{z} k_2}{e^{\overline{z}k_2/V\eta_u}-1}   \right]-\frac {\overline{z}}2(k_0+k_1),
\label{solucion}
\end{equation}
and it is bounded by the two limiting values  
\begin{eqnarray}
\sigma(V\to 0)&=&f^{th}-\frac {\overline{z}}2(k_0+k_1)\\
\sigma(V\to \infty)&=&f^{th}-\frac {\overline{z}}2(k_0+k_1+k_2)
\label{limites}
\end{eqnarray}
 Note that :
 \begin{itemize}
 \item  The friction force decreases as the driving velocity increases, an effect called {\em  velocity weakening}, and observed in tribology experiments for different materials, especially at very low velocities \cite{Kilgore1993, Capozza2013}
 \item The  velocity weakening displays a characteristic $1/V$ decay at large $V$. 
 \item The dependence on $k_0$ and $k_1$ is limited to the last term in Eq. (\ref{solucion}), that accounts for a constant shift of the whole $\sigma(V)$ curve.
 \end{itemize}
 Velocity weakening
 is a necessary link between the static ($V=0$) and kinetic ($V>0$) friction coefficients, and as such is known to be crucial 
 in the triggering of instabilities in sliding systems, leading to stick-slip motion \cite{Baumberger1995} and to the existence of earthquakes in sliding tectonic faults \cite{Scholz2002a}.
For a review on nanoscale models of friction and experimental results on nano-tribology one should consult \cite{Vanossi2013} or the letter \cite{Krim2002}, which contains accessible references to the literature.
In our model, velocity weakening is a direct consequence of the viscoelastic relaxation. 
In fact, the model without viscoelasticity lacks any velocity dependence of the friction force, as in that case there is no internal time scale to compete with the driving time scale. 
It should be noted that the most commonly observed velocity-weakening friction law is only logarithmic in $V$, but cannot be expected here because we introduced a single relaxation time scale, in a mean field model.
This echoes the results found in \cite{Thøgersen2014}, where a $1/V$ velocity weakening law is found to occur in a non interacting block model with random friction coefficients, where the relaxation time scale is present thanks to the non-instantaneous slips.
One may also note that we do not expect any of the velocity-strengthening scenarios (as those proposed in \cite{Bar-Sinai2014}) to occur either, since we do not account for any of the faster mechanisms that become relevant under fast driving.

 We now turn on the numerical study of the mean field version of  Eqs.~(\ref{Eq:fast1}) and (\ref{Eq:relaxation1}) using an uncorrelated distribution of pinning wells, with mean $\bar z$, namely $g(z)=\bar z ^{-1}e^{-z/\bar z}$. In practice we implement the Fokker-Planck method used in  \cite{Jagla2014a}, adapted to the case of finite driving velocity. All technical details are left to Appendix \ref{app:numerical_master_equation}.

\begin{figure}
\includegraphics[width=0.48 \textwidth]{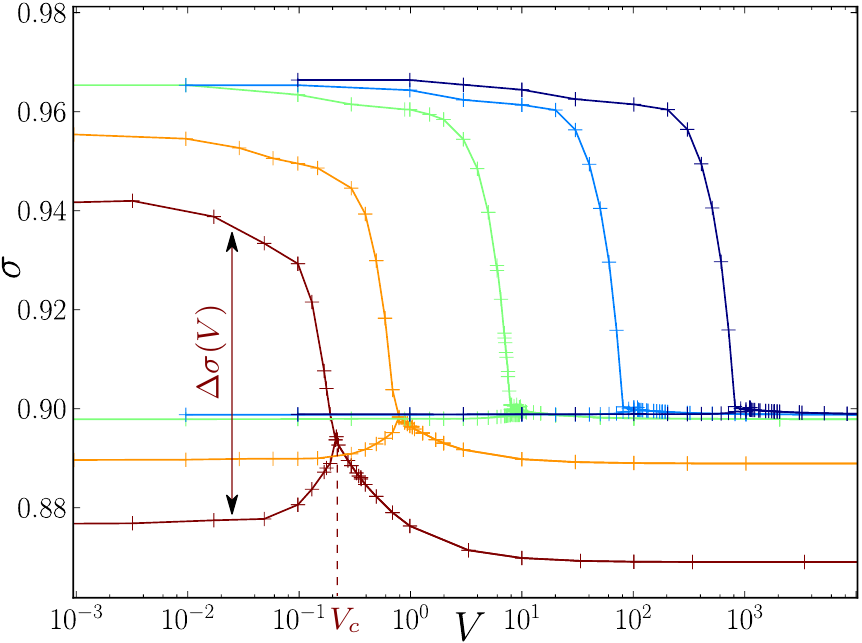}
\caption{
(Color online)
\textit{Friction force against driving velocity}, for different values of $k_0$ ($k_0=0.3, 0.1,0.01, 0.001, 0.0001$, from left to right); and constant $k_1=0.1$, $k_2=0.9$, $\overline{z}=0.1$.
In the slow driving regime (on the left), motion occurs mostly through the abrupt slips during which stress falls from the stress Before slip, $\sigma_B$ (up), to the stress After slip, $\sigma_A$ (down).
In the faster driving regime, the dynamics is stationary and friction decreases with velocity.  
Crosses indicate the values of $V$ for which the numerical integration has actually been performed.
\label{Fig:velocityWeakening1}
}
\end{figure}

In Fig. \ref{Fig:SigmaVSt}, the evolution of the stress over time is compared for three values of the velocity: at slow driving velocities the stress $\sigma(t)$ oscillates periodically with an amplitude denoted $\Delta \sigma (V)$,
 while at fast driving a stationary value is reached (i.e.~$\Delta \sigma (V)=0$).
Note that the amplitude $\Delta \sigma(V)$ of the oscillations is also the width of the stress drops or ``gaps'' that occur during the system-size events.

\begin{figure}
\includegraphics[width=0.48 \textwidth]{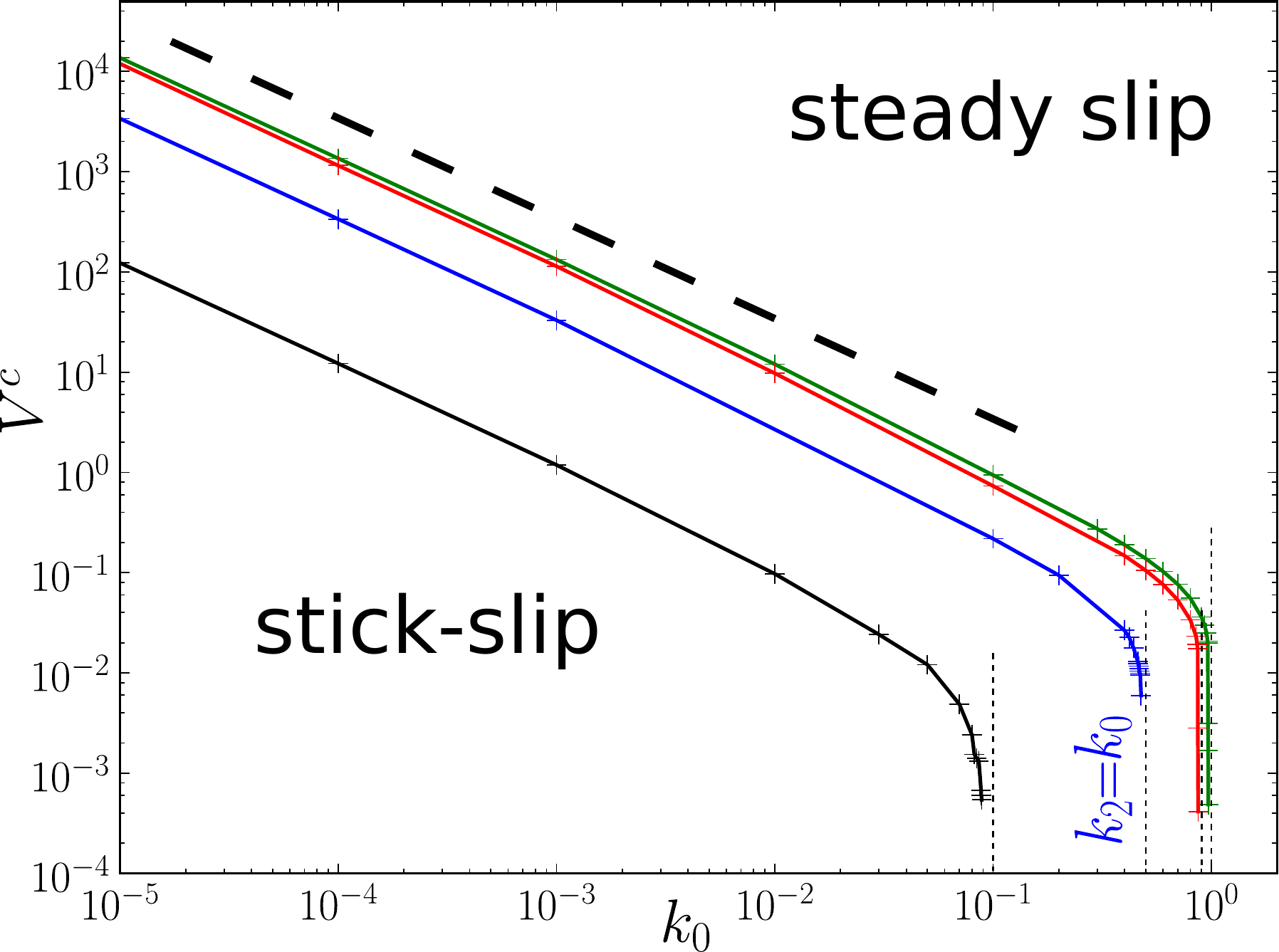}
\caption{
(Color online)
\textit{Phase diagram in the $k_0-V$ plane}, for different sets of $(k_1,k_2)$ values: from bottom to top we used $(k_1=0.0,k_2=1.0), (0.1,0.9), (0.5,0.5), (0.9,0.1)$. In the lower-left region, the system behaves in a non-stationary way, i.e.~we have stick-slip motion. 
For $k_0 \geq k_2$, the system is never non-stationary, even at quasistatic driving, as predicted in \cite{Landes2014}. 
The bold dashed line has slope one, indicating a scaling of the critical velocity as $V^c \sim k_0 ^{-1} $ for small $k_0$'s.
The thin dashed vertical lines indicate the asymptotic behavior $V^c\to 0$ when $k_0 \to k_2^-$.
\label{Fig:phaseDiagram}
}
\end{figure}

This behavior points to a bifurcation of the dynamics as velocity is reduced. 
To study this effect, we report in Fig. \ref{Fig:velocityWeakening1} the maximum and minimum of the friction force over time, as a function of the driving velocity $V$, for various values of $k_0$.
 We observe that the stress gap 
 vanishes smoothly at the transition point $V=V^c$, pointing to a ``second order'' dynamical phase transition in the order parameter $\Delta\sigma$. 
 In Fig. \ref{Fig:phaseDiagram} we show the full phase diagram  of the system in the $k_0$-$V$ plane. There, for various sets of $(k_1,k_2)$ we observe a divergence of $V^c$ as $1/k_0$.
This $1/k_0$ dependence at low $k_0$ comes directly from the uniform increase of the pulling force as $k_0 V t$ and implies that for any non zero values of $k_1, k_2$ and $V$, there will always be a value of $k_0$ below which some stick-slip dynamics occurs, even at very large velocities. 

\begin{figure}
\includegraphics[width=0.48 \textwidth]{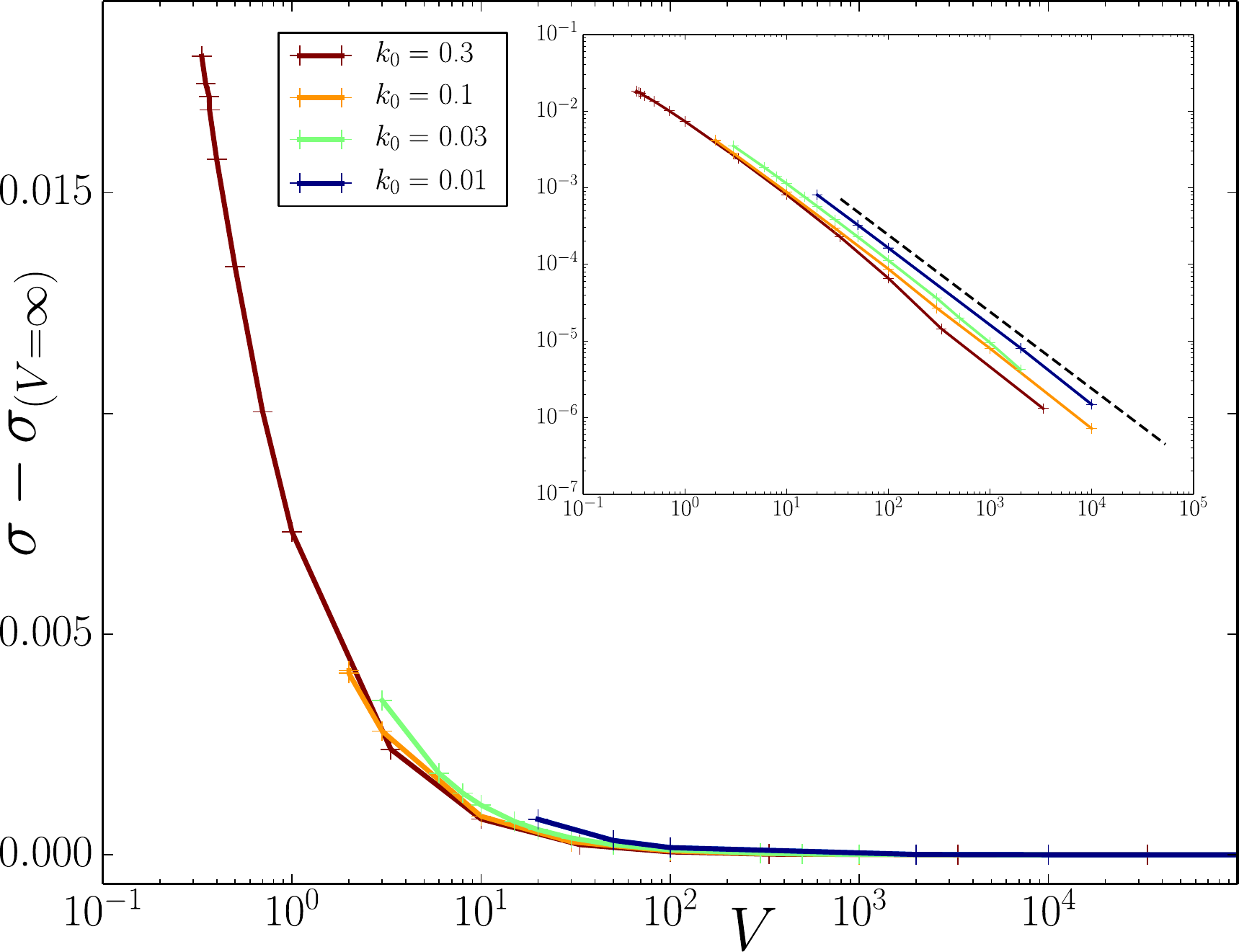}
\caption{
(Color online)
\textit{Velocity weakening in the steady slip regime} ($V>V^c$), far from the critical point.
We observe that $\sigma$ decreases when the velocity increases. For $V \to \infty$, the decay in the friction force towards the limiting large-velocity value $\sigma_{V=\infty}$ goes as $\sim V^{-1}$. 
Inset: plot in the log-log coordinates.  The dashed line gives the pure power law with exponent $-1$. 
We used the same color code as in previous figures (curves that go to larger values are the lower $k_0$'s).
\label{Fig:Asymptotic_Sigma_Right}
}
\end{figure}

In the stationary regime, our simulations confirm,  overall, the friction behavior of Eq.~(\ref{solucion}). In particular, in  Fig.~\ref{Fig:Asymptotic_Sigma_Right} we observe a clear velocity weakening with a characteristic  $1/V$ decay towards $\sigma(V\to \infty) =1-(k_0+k_1+k_2)\bar{z}/2$.

\section{Slide-hold-slide Protocol}
\label{sec:slide_hold_slide}

Slide-hold-slide experiments are an important tool in the investigation of the tribology of solids. 
When the sliding of a solid is interrupted for some time $\Delta t$, the contacts at the surface of the solid can strengthen over time, so that when sliding is resumed, a peak in the friction force has to be overcome before one recovers the stationary friction force. 
The amplitude of the friction peak increases with the hold time $\Delta t$ since the relaxation is more effective when it has more time to act. 
Our model has all the necessary ingredients to reproduce the peak in the friction force after 
a hold period. In fact, when driven at a finite velocity, the viscoelastic elements do not have the time to
completely relax to the most convenient (i.e., lowest energy) configuration at each global position. If a hold time is given to the system, the mechanical energy reduces as the viscoelastic elements relax. Upon resuming driving, this lower energy configuration requires a larger stress to initiate motion again. Thus the effect is expected to become stronger as the hold time increases, saturating at hold times much larger than the viscoelastic relaxation time.

\begin{figure*}
\includegraphics[width=0.46 \textwidth]{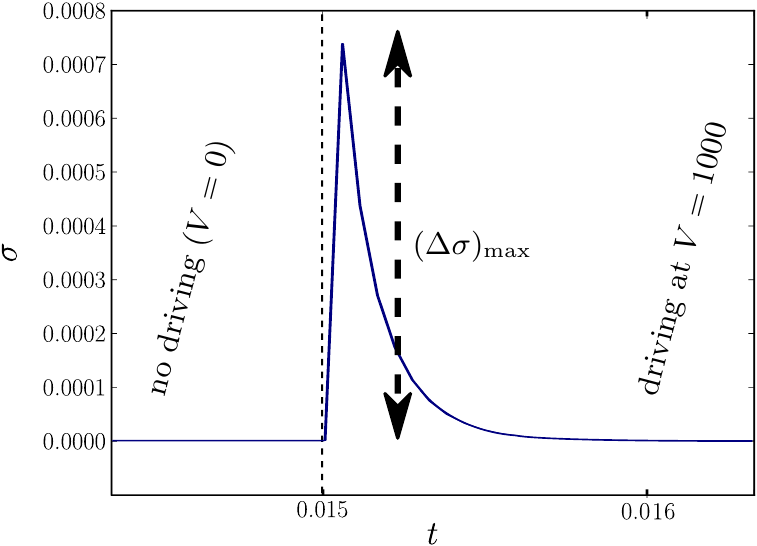}
\hfill
\includegraphics[width=0.48 \textwidth]{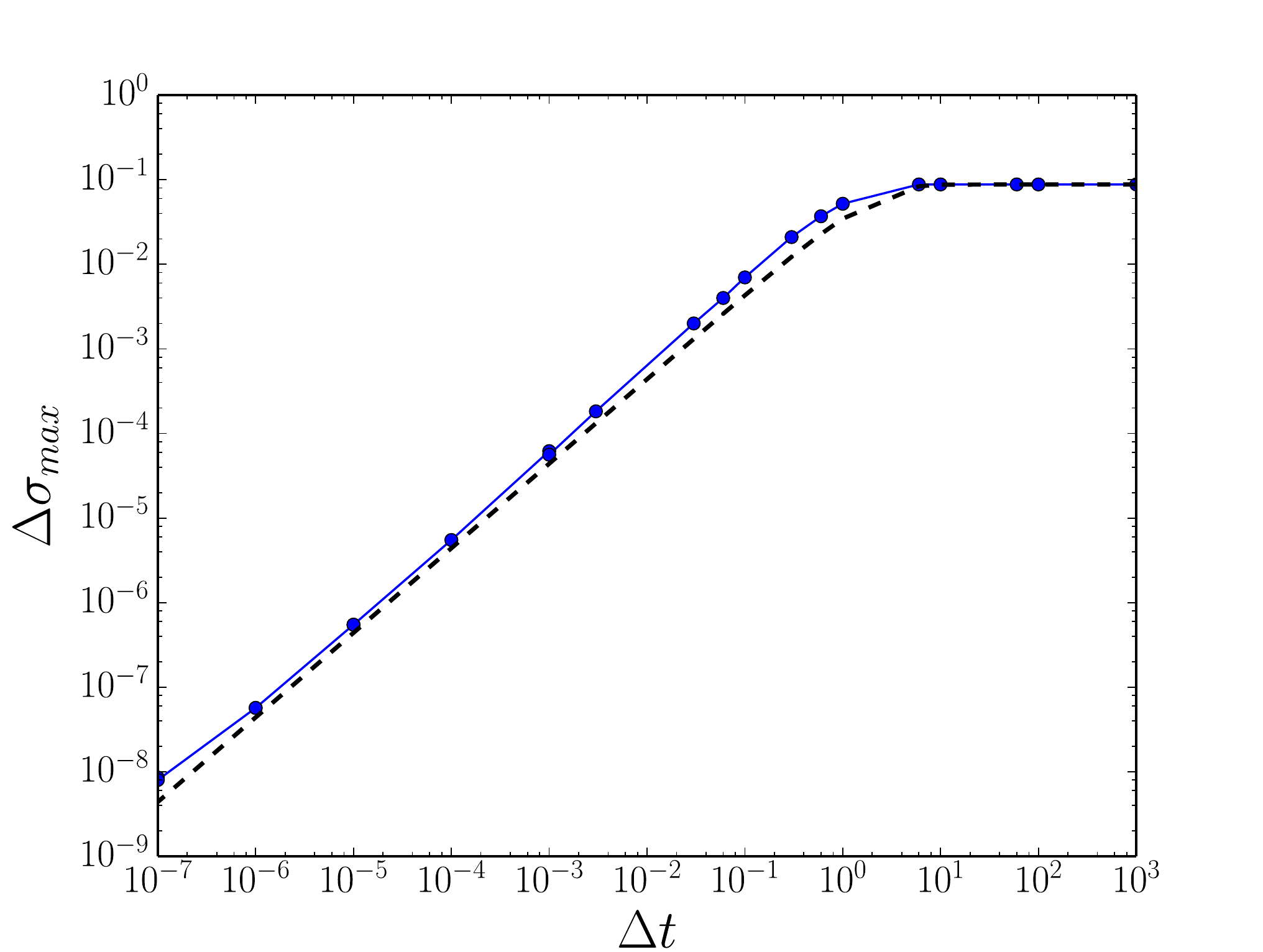}
\caption{
(Color online)
\textit{Slide-hold-slide experiment.} Left: Evolution of the stress $\sigma$ over time under this protocol, around the time when driving is resumed ($t=0.015$ here).
The driving velocity is $V=1000$.
\label{Fig:Overshoot_VS_DeltaT_1}\\
Right: For many different hold times $\Delta t$, we record the maximal deviation $(\Delta \sigma)_{max}$ from stationary stress (blue points). This deviation corresponds to the gap between static and kinetic friction coefficient.
In our model we observe a linear growth of the gap with $\Delta t$ while in experiments a logarithmic increase is measured. 
The dashed line gives the purely exponential saturation: $\Delta\sigma_{\text{max}} = \Delta\sigma_{\infty} (1-e^{-\Delta t/\tau}) $, with $\tau=2$ and $\Delta\sigma_{\infty}\equiv \Delta\sigma_{\text{max}}(\Delta t =\infty)$.
\label{Fig:Overshoot_VS_DeltaT_2}
}
\end{figure*}

To simulate this process, we set the driving velocity $V$ to a value at which a steady sliding is observed. 
Once we reached a stationary dynamics, the driving is stopped for some time interval $\Delta t$ and then resumed. 
The evolution of the friction force during this protocol in our model is shown in 
Fig. \ref{Fig:Overshoot_VS_DeltaT_1} (left). 
The height of the friction peak as a function of the hold time $\Delta t$ is shown in Fig. \ref{Fig:Overshoot_VS_DeltaT_2} (right). 

We observe some similarities and some differences when comparing our results with experimental observations.
 In experiments, during the hold time the stress shows a relaxation towards some lower value, something which is not observed in our model. 
 The reason is that stress relaxation can occur only if some secondary avalanches (aftershocks) are triggered by the viscoelastic relaxation. 
 When the pinning thresholds $f^{th}$ have a single value, then aftershocks are excluded and the decrease can not happen. However in models with a wide distribution of pinning thresholds this relaxation has been observed \cite{Jagla2010a} and is compatible with experiments. 

 Qualitatively, the friction peak is similar in our simulations and in experiments: in both cases its height increases as a function of the hold time, but in experiments the increase is usually reported to be logarithmic, while here we observe an exponential  saturation at large 	 $\Delta t$.
 The reason for this difference is that our model contains only a single relaxation time constant, which naturally defines a typical time scale, above which the system is fully relaxed and thus no longer evolves.

\section{Analysis in terms of rate-and-state equations}
\label{sec:RSF_analysis}

It has been long realized that the friction force cannot be described by a single valued function of the instantaneous velocity. The history  of the contact plays an important role in determining the actual friction force. Our model is an example of such a case, since the value of the friction force depends on the state of the viscoelastic elements, which in turn depend on the history of the system. We thus briefly recall the standard rate and state formalism, and then show how it provides an appropriate framework to understand our results.

\subsection{The RS Formalism}

Phenomenologically, the behavior of frictional contacts has been successfully described through the formalism named rate-and-state (RS) friction, originated in the works of Dieterich and Ruina \cite{Dieterich1979, Ruina1983}. 
Instead of assuming that there exists only a kinetic and a static friction coefficient, the RS formalism assumes that the friction coefficient continuously depends on the relative velocity $v$ between the two surfaces (the rate variable) and on a state variable usually called $\theta$.
We recall that by definition, the friction coefficient $\mu$ acts as a threshold for the friction force $\sigma$ actually arising from the contacts: we always have  $\sigma\leq \mu F_N$, where $F_N$ is the force  normally applied on the solid. 
The usual RS form for $\mu(v,\theta)$ is:
\begin{equation}
\mu(v,\theta) = \mu_0+ a\log(v)+b\log(\theta/D_c)
\label{rs}
\end{equation}
where $a, b$ are positive constants. 
Note that we use a small $v$ to indicate the instantaneous velocity, which may differ from the driving velocity $V$. 
The $a$ term describes the so-called ``direct effect'', 
 the increase of friction with increase of the relative velocity commonly observed in many materials. 
 The state variable $\theta$ is supposed to follow a second equation, usually written: 
\begin{equation}
\dot \theta= 1-\frac{\theta v}{D_c},
\label{2}
\end{equation}
where $D_c$ is the ``critical slip distance'', i.e. the amount of slip (of the center of mass of the sliding block) necessary to break a newly formed junction. 

Under steady sliding, we have $\dot \theta=0$, $\theta=D_c/v$, and the RS equations (\ref{rs}), (\ref{2}) simplify into:
\begin{equation}
\mu(v,\theta) =\mu_0+(a-b)\log(v).
\label{rs2}
\end{equation}
If $b>a$ this equation describes the phenomenon of velocity weakening, namely a reduction of the friction coefficient when velocity increases. Velocity weakening is a crucial ingredient involved in the description of seismic phenomena\cite{Scholz2002a}. 
In the case in which the contact is at rest ($v=0$), we get $\dot \theta=1$, i.e., $ \theta(t)=\theta_0+t$, and according to (\ref{rs}) we obtain an increase of the static friction coefficient with the time of contact.
RS equations have been used to describe the behaviors of frictional systems under a variety of non-steady sliding conditions, providing an excellent phenomenological description for numerous systems.

\subsection{Interpretation of our model in terms of RS Equations}

Under uniform driving, our model displays two features that are common to many frictional systems: a velocity weakening friction law, and a bifurcation to a stick-slip regime for low driving velocities. We have described in Section \ref{sec:constant_drive_proto} how the velocity weakening effect appears in the model. Now we will see that assuming the existence of this effect, the bifurcation to a stick-slip regime at low velocities is captured in all its detail by the use of the RS formalism.

The usually assumed form Eq. (\ref{rs2}) of the friction coefficient is not appropriate to accurately describe the results observed in our model in the steady sliding regime. On the one hand, our model does not have any ingredient that could produce a direct effect in the RS equation (\ref{rs}), suggesting that we should use $a=0$. On the other hand, the logarithmic $b$ term is not appropriate in our case, mainly because we have considered only a relaxation mechanism that is described by a single time constant, the one associated to the dashpot and $k_2$ springs.
Instead, from our simulations of Sec. \ref{sec:constant_drive_proto} (as for instance those in the right part of Fig. \ref{Fig:velocityWeakening1}) 
 we can construct a simple analytical expression for the friction coefficient in the steady sliding regime:
\begin{equation}
\mu^{\text{steady}} \simeq f^{th}-\frac{\bar z}{2}(k_0+k_1+k_2) + \frac{.007}{v+.078}
\label{g}
\end{equation}
where the numerical values correspond to $k_1=0.1$, $k_2=0.9$ and $\overline{z}=0.1$.
We remark that this is just a simple, although accurate fitting to the numerical results that allows the analytical treatment presented below, but it has no additional particular significance.

We now assume that under any non-steady sliding situation, the
 value of the friction coefficient can still be written in terms of the function found in the steady state. Using the fact that in the steady state $v\equiv D_c/\theta$, we write:
\begin{equation}
\mu(\theta) = f^{th}-\frac{\bar z}{2}(k_0+k_1+k_2) + \frac{.007}{D_c/\theta+.078},
\label{1}
\end{equation}
The time evolution of $\theta$ will be assumed to be described by the standard RS evolution, Eq. (\ref{2}).
The sliding contact is coupled to a driving spring pulled at constant velocity $V$, so that the pulling force reads:
\begin{equation}
\sigma=(Vt-x)k_0
\label{3}
\end{equation}
where $x$ is the spatial average coordinate of the contact, i.e. $\dot x=v$. 
This force $\sigma$ is also the actual instantaneous friction force arising from the contacts. 
Our aim is now to determine the temporal evolution of $x$ and $\sigma$ and reproduce the bifurcation behavior that we obtained in Section \ref{sec:constant_drive_proto}.

\begin{figure}
\includegraphics[width=0.48 \textwidth]{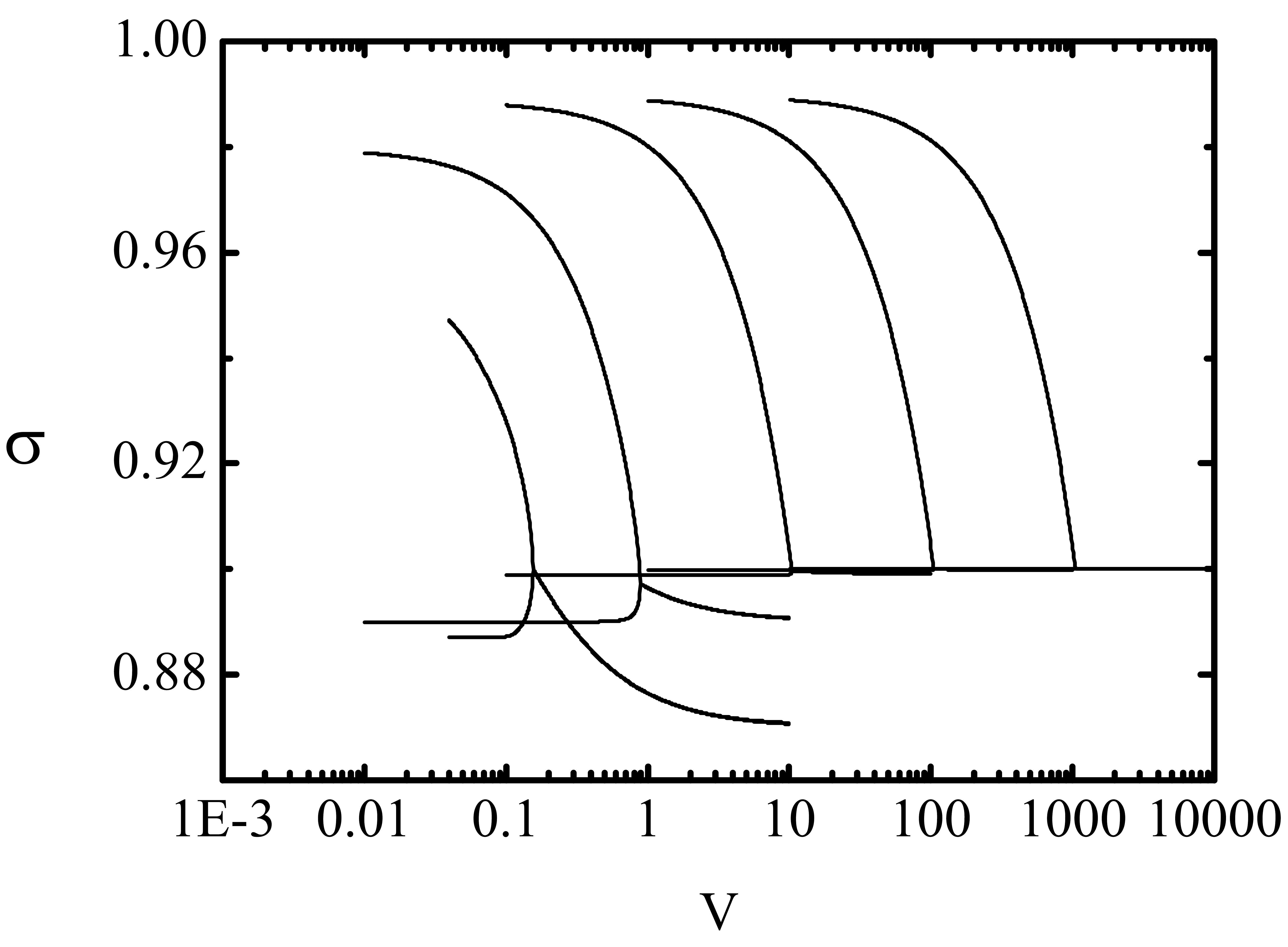}
\caption{
Friction force as a function of velocity for the same set of values as in Fig. \ref{Fig:velocityWeakening1}, using the RS formalism. The form of the steady sliding part of the curves is introduced by hand. 
We used a value $D_c=0.06$ to achieve the best fitting.
The RS formalism yields the form of $\sigma_{max}$ and $\sigma_{min}$ for $V<V^c$. 
}
\label{fig:rs1}
\end{figure}

If steady sliding is assumed, the force $\mu F_N$ must balance the pulling force, and we get $\theta=D_c/V$, $v=V$.
However, steady sliding may be unstable due to the following process: if at some moment $x$ is stopped, the pulling force starts to increase according to $\sigma=\sigma^0+Vtk_0$. On the other hand, the friction force starts to increase due to the increasing of $\theta$, according to $\mu=\mu(\theta_0+t)$. If the initial increase of $\mu F_N$ is more rapid than the increase of $\sigma$, the contact will stay stuck as long as $\mu F_N>\sigma$, until eventually the pulling force becomes larger than the pinning force, and a rapid slip moves the value of $x$ to a new position. 
If $\mu F_N<\sigma$, slip occurs and is essentially instantaneous since we have not added inertia to the contact.
During the slip stage, the value of $\theta$ is reduced following Eq. (\ref{2}), that in the rapid slip limit can be written as $\d \theta=-\theta \d x/D_c$ and analytically integrated. Slip finishes when $\sigma$ drops below $\mu F_N$ again, and a new stick slip cycle begins. 
We have integrated Eqs. (\ref{1}), (\ref{2}), (\ref{3}) and in fact obtained steady sliding for large $V$, and a stick slip behavior at small $V$, in which the value of $\sigma$ oscillates between two values $\sigma_{\max}$ and $\sigma_{\min}$. 
In Fig. \ref{fig:rs1} we plot the values of $\sigma_{\max}$ and $\sigma_{\min}$ (or the single value $\sigma$ in case of steady sliding) as a function of $V$ for the same set of $k_0$ values used in Fig. \ref{Fig:velocityWeakening1}.
It can be seen that the value of $D_c$ enters in the problem only in combination with the spring stiffness, as $kD_c$. In Fig. \ref{fig:rs1} we used $D_c=0.06$ to achieve the best fitting with the results of Fig.  \ref{Fig:velocityWeakening1}. We note that as suggested by its original meaning, $D_c$ is of the order of $\overline{z}$, which is the distance at which the correlations between pinning forces disappears.
 In addition to the coincidence of the two figures in the steady sliding regime (which is enforced by hand through the choice of the $\mu^{\text{steady}}$ function in Eq. (\ref{g})), we see that the RS formalism gives a very good coincidence in the low velocity, stick-slip regime.

A careful analysis of our RS equations near the bifurcation point shows that the two values 
$\sigma_{max}$ and $\sigma_{min}$ depart symmetrically from the branch of steady sliding, and are such that $\sigma_{max}-\sigma_{min}\sim (V^c-V)^{1/2}$, which
corresponds to a Hopf bifurcation. Note however that the validity range of these scaling becomes progressively smaller as $k_0$ is reduced, and in the $k_0\to 0$ limit we get the scaling $\sigma_{\min}\sim\text{ const}$, $\sigma_{\max}-\sigma_{\min}\sim (V^c-V)$.

\section{Conclusion}
\label{sec:conclu}

We have presented a detailed analytical and numerical analysis of a viscoelastic model of friction in mean field approximation, in which the driving velocity competes with the time scale of the viscoelastic effects within the system. 

Our main findings are the following.
At low driving velocities, we obtain a stick-slip dynamics with amplitudes of the stress oscillations 
that decrease with increasing driving velocity ($V$) or increasing driving stiffness ($k_0$).
Beyond a certain critical driving velocity ($V>V^c$), the amplitude of these oscillations becomes zero, i.e. the sliding is smooth.
The transition between stick-slip and smooth sliding occurs in a continuous manner and is very well described by a phenomenological rate-and-state analysis.
In the smooth sliding regime the friction force reduces as a function of the driving velocity, reproducing the well known velocity-weakening phenomenology.
In our model, this effect is originated in the existence of viscoelastic elements that set an additional time scale for the dynamics. 
Finally, the response of our model to intermittent driving allowed us to reproduce qualitatively an important aspect of the aging of contacts, namely the increase of the static friction with time of contact at rest.
Overall, we believe our model reproduces many well-known features of real tribology, and gives a well defined model on which many assumptions and predictions of phenomenological theories (like rate-and-state equations) can be investigated in depth.

\section*{Acknowledgements}

EAJ is financially supported by CONICET (Argentina). Partial support through grant PICT 2012-3032 (ANPCyT, Argentina) is also acknowledged.
We acknowledge the support by the France-Argentina MINCYT-ECOS A12E05.

\appendix

\section{Detailed Derivation of the Equations of Motion}
\label{sec:pedestrian}

\begin{figure}
\includegraphics[scale=0.9]{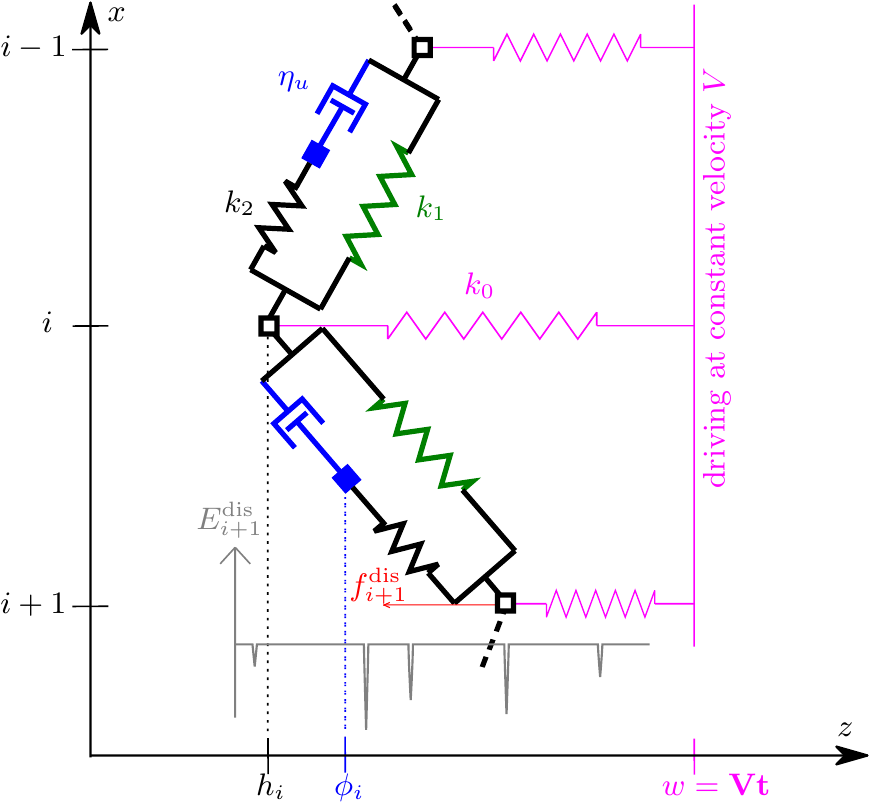}
\caption{
\textit{Sketch of the one-dimensional viscoelastic interface model.}
The interface itself (bold black line) consists in blocks set at discrete lattice sites $i,i+1, \dots$ along the $x$ axis (in higher dimensions, $x\in \mathbb{R}^ d$). 
The blocks can evolve along the $z$ axis: they are identified by their locations $h_i,h_{i+1},\dots$ (empty squares) and bound together via a combination of springs ($k_1, k_2$) and dashpots ($\eta_u$).
The viscoelastic interaction introduces an internal degree of freedom $\phi_i$, represented by full squares (blue).
Driving is performed via springs $k_0$ linked to 
the 
position $w=Vt$ (thin purple lines). 
The disorder force $f^\text{dis}_i$ (red) for the site $i$ derives from a disordered energy potential $E^\text{dis}_i(z)$ (grey).
\label{Fig:model_detailed}
}
\end{figure}

We first study the one-dimensional case, as presented in Fig. \ref{Fig:model_detailed}.

The interface is decomposed in blocks of mass $m$, labelled $i$ and moving along horizontal rails $h_i$. 
The action of the dashpot is to resist the change in $\phi_i - h_i$ via viscous friction, with a resulting force on $h_i$ given by $\eta_u  \partial_t (\phi_i-h_i)$.
The blocks move in a medium with some effective viscosity $\eta$ and we study the overdamped regime, $m \partial_t^2 h_i \ll \eta\partial_t h_i$. 
As each block is described by two degrees of freedom $h_i$ and $\phi_i$, the time evolution is governed by two equations.
We now provide a pedestrian derivation of the equations, for the sake of completeness. 
The first equation comes from the force balance on $h_i$:
\begin{align}
\eta \partial_t h_i 
=&  f^\text{dis}_i(h_i) + k_0 (w-h_i) + k_1(h_{i+1} - h_i)  \notag  \\
&+ k_1 (h_{i-1} -h_i) + \eta_u \partial_t(\phi_i -h_i) + k_2 (\phi_{i-1}-h_i)
\label{Eq:viscoqEW1}
\end{align}
The second equation is derived from the force balance on $\phi_i$: 
\begin{align}
0= k_2(h_{i+1} - \phi_i) + \eta_u \partial_t(h_i - \phi_i) 
\label{Eq:NewtonOnPhi}
\end{align}
where we assume that the internal degree of freedom $\phi_i$ has no mass.
Similarly the force balance on $\phi_{i-1}$ yields:
\begin{align}
0= k_2(h_{i} - \phi_{i-1}) + \eta_u \partial_t(h_{i-1} - \phi_{i-1}) .
\label{Eq:NewtonOnPhi-1}
\end{align}
In order to let the Laplacian term $k_2 (h_{i+1}- 2h_i+h_{i-1})$ appear, we introduce the variable 
\begin{align}
u_i \equiv \phi_i - h_i + h_{i-1} - \phi_{i-1},
\end{align}
which represents the elongation of the dashpot elements connected to site $i$.
We inject \req{NewtonOnPhi} into \req{viscoqEW1} to get rid of the time derivatives, and we subtract \req{NewtonOnPhi-1} from \req{NewtonOnPhi} to obtain \req{detaille2}:
\begin{align}
\eta \partial_t h_i =& f^\text{dis}_i(h_i) + k_0 (w-h_i) \notag\\
 &+ (k_1+k_2)(h_{i+1} -2 h_i + h_{i-1} ) 
- k_2 u_i  
\label{Eq:detaille1}\\
\eta_u \partial_t u_i 
=& k_2 (h_{i+1}- 2h_i+h_{i-1}) - k_2 u_i. 
 \label{Eq:detaille2}
\end{align}

A more elegant notation using the Laplacian operator $\nabla^ 2$ is:
\begin{align}
\eta \partial_t h_i &= f^\text{dis}_i(h_i) + k_0 (w-h_i) + k_1 \nabla^ 2_i h_i  +k_2 (\nabla^ 2_i h_i  -u_i) \notag \\
\eta_u \partial_t u_i &= k_2 (\nabla^ 2_i h_i  -u_i). \label{Eq:viscoDetaille3}
\end{align}

To generalize this to higher dimensions (on a square lattice), one simply has to connect each block $h_i$ to its neighbors via viscoelastic elements, using a single orientation per direction.
The equations obtained are exactly \req{viscoDetaille3} if we reinterpret the label $i$ as referring to $d$-dimensional space, the Laplacian $\nabla^ 2$ as the $d$-dimensional one, and the $u_i$ variable as:
\begin{align}
u_i = \sum_{j=1}^ {d}( \phi_j - h_j) + \sum_{j'=d+1 } ^ {2d} (h_{j'} - \phi_{j'}) \label{Eq:definitionGeneraleDe_u},
\end{align}
where indices $j$ denote the $d$ first neighbors, connected via a dashpot followed by the spring $k_2$ (and $k_1$ in parallel) and indices $j'$ denote the last $d$ neighbors, connected via the spring $k_2$ followed by a dashpot  (and $k_1$ in parallel).

We study the mean field limit via the fully connected approximation.
In practice, each block position $h_i$ interacts with the positions of all other blocks via $N-1$ springs of elastic constant $k_1/N$ ($N$ being the number of blocks in the system) and via $N-1$ viscoelastic elements (i.e.~spring in series with a dashpot). 
As usual for fully connected systems, the final equation for the site $i$ is obtained by replacing any occurrence of $\Delta h_i$ with $\overline{h}-h_i$.

\section{The Numerical Methods}
\label{app:numerical_master_equation}

In previous works \cite{Jagla2014a, Landes2014} we showed how to translate the stochastic dynamics into a master equation (or Fokker-Planck equation), but only in the case in which $V\to 0$.  Here we generalize the procedure to any velocity $V$. Our Cython code is fully available \cite{CodeLandes2014}.

It is useful to describe the dynamics using the local variable,  $\delta_i$:
\begin{align}
\delta_i \equiv 1 - k_0 (w-h_i)  - (k_1+k_2) (\overline{h}-h_i) + k_2 u_i,
\label{Eq:deltas_depinning_MF}
\end{align} 
which represents the amount of additional stress that a site can hold before becoming unstable: 
as long as $\delta_i >0, \forall i$, the block is stable and its position, $h_i$, remains fixed when $\delta_i=0$ the block jumps to the next pinning well at a random distance $z$: $h_i \to h_i+z$.
To define the dynamics in terms of $\delta$ variables, we need to split it in a {\em fast} part $\delta^F$, and a {\em relaxed} one $\delta^R$: 
\begin{align}
 \delta^F _i &= 1 - k_0 (w-h_i) - (k_1 + k_2) (\overline{h}-h_i)\notag  \\
  \delta^R _i &=   k_2 u_i,
  \label{Eq:defdeltaF} 
\end{align}
such that $  \delta_i = \delta^F _i + \delta^R_i $.
The blocks dynamics is controlled by three processes:
\begin{itemize}
\item[(i)] \textit{The avalanches.} 
When a block is unstable ($\delta_i \leq 0$), it moves to the next pinning wells: $\delta_i^F \mapsto \delta_i^F+ z(k_0+k_1)$, with $z$ drawn from $g(z)$.
Each jump $z$ is followed by a stress redistribution $\delta_j^F \mapsto \delta_j^F - z k_1/N$.
These drops can trigger other instabilities.
The characteristic duration of such an avalanche is $\tau_0 = \eta_0/ \max(k_0,k_1,k_2)$.
\item[(ii)] \textit{The driving.} 
External driving increases over time: for instance $w=Vt$ in the uniform driving protocol. 
In terms of $\delta$, driving means that over a time step $\d t$, $\delta_i^F \mapsto \delta_i^F - k_0 V \d t$. 
This driving can happen until a new instability is triggered (step (i)).
The characteristic time scale of driving is $\tau_D = \overline{z}/V$.
 \item[(iii)] \textit{The relaxation.} 
In absence of instabilities, the $h_i$'s are constant, and Eq.~\ref{Eq:relaxation1} reduces to:
\begin{align}
\delta^R_i(t)= k_2 (\overline{h}-h_i)+ \lp \delta^R_i(t_0) - k_2 (\overline{h}-h_i) \rp e^{-k_2\frac{t-t_0}{\eta_u}}
\end{align}
where $t_0$ is the time at which the last avalanche occurred. 
Note that $h_i$ does not evolve during relaxation or driving, so that relaxation can happen until a new instability is triggered (step (i)).
The characteristic time scale of relaxation is $\tau_u = \eta_u / k_2 $.
\end{itemize}
In this paper we study the case $\tau_0 \ll \tau_u \sim \tau_D$ and focus on the case of the mean field model where fluctuations vanish and the description of the system via a simple probability distribution becomes exact.
Indeed, the sole distribution $P(\delta)$ does not provide enough information to fully characterize the system and its evolution.
We consider the joint probability density distribution $P( \delta^F, \delta^R)$. 
The quantity $P( \delta^F, \delta^R,t) \d \delta ^ F \d \delta^ R$ represents the probability for a site drawn at random to have a particular set of values of $\delta^F, \delta^R$ and can be computed numerically  starting from the dynamical rules that apply to the $\delta$'s.

To be concrete, we discretize $P(\delta^F, \delta^R)$ with a bin $\varepsilon$.
The distribution probability is then a matrix $P_{i,j}$ where $P(\delta^F = \varepsilon i, \delta^R = \varepsilon j) \d \delta^ F\d \delta^ R= P_{i,j}$. 
 We use a time step $\d t = \varepsilon / k_0 V$ and define the constant  $\kappa=k_0+k_1+k_2$. 

The finite velocity is expressed through the fact that every time there is some driving of $w$ by a quantity $\d w=V \d t$, there is also relaxation during a time $\d t$. In particular, we define the relaxation factor $R(\d t) =1-e^{-\d t k_2 / \eta _u}$. For the avalanches two crucial quantities should be defined: (1) the fraction of unstable sites  $P_{\text{unst}}^\text{tot} \equiv \varepsilon \sum_{i,j|i+j<0} P_{i,j} = \varepsilon \sum_{j} \sum_{i'|(i'+j<0)} P_{i',j}   $ (2) the stress redistribution $P_{\text{unst}}^\text{tot} \overline{z}(k_1+k_2)$ that follows the stabilization of the unstable sites.
When  $P_{\text{unst}}^\text{tot} \overline{z}(k_1+k_2)>1$, the avalanche increases geometrically over the time steps, this is why it is practical to define a ``critical'' value 
$P_0^ c= \frac{1}{\overline{z}(k_1+k_2)}.$

The sketch of the algorithm is the following. 
\begin{itemize}

\item {\em Relaxation process}:
\begin{itemize}
\item 
Compute $j_{\infty}(i) $, the bin associated to the fully relaxed state, $ \delta^R_{i,\infty}= \frac{k_2}{\kappa}(\overline{\delta^ F}-\delta^ F)$. 
\begin{align}
j_{\infty}(i)=  \text{Int} \left(  k_2  \frac{- i +  \sum_{i',j} i' P(i',j)  }{\kappa} \right). \nonumber
\end{align}
\item Relaxation corresponds to shift \footnote{Instead of crudely taking the integer part of $ (j_{\infty}(i) - j) R(\d t)$, it is numerically much more stable to split the  shift over  the two  bins, using a linear interpolation} $P_{i,j}$ to $P_{i,j+\text{Shift}}$ (where $\text{Shift} = \text{Int} \left[  (j_{\infty}(i) - j) R(\d t) \right]$), set $r=1$ and perform the {\em Avalanche process}.
\end{itemize}

\item {\em Avalanche process}: consists of driving and jumps. 
\begin{itemize}
\item Driving:
\begin{align}
P_{i,j}\leftarrow P_{i,j} + \left( P_{i+1,j}-P_{i,j} \right) r \label{AvaProcess}
\end{align}
\item Compute  $P_{\text{unst}}(j)=\sum_{i'|i'+j<0} P_{i',j} $
\item Jumps:
\begin{align}
P_{i,j}\leftarrow P_{i,j} +\frac{\varepsilon}{\kappa}    g\left(\frac{\varepsilon (i + j)}{\kappa}\right)   P_{\text{unst}}(j)
\end{align}
\begin{align}
P_{i,j} &\leftarrow 0 \quad \text{if} \; i+j<0 \nonumber
\end{align}
\item Compute $P_0=\sum_{i=-j} P_{i,j}$ 
\item If $P_0 \geq P_0^ c$, set $r=1$ and perform the {\em Avalanche process} again. 
\item Else
\begin{itemize}
\item Compute $P_{\text{unst}}^{\text{tot}} = \sum_j P_{\text{unst}}(j)$
\item If $P_{\text{unst}}^{\text{tot}} \geq P_0^ c /100$, set $r=\min(1, P_{\text{unst}}^{\text{tot}} / P_0^ c)$ and perform the {\em Avalanche process} again.
\item 
Else, 
perform the  {\em Relaxation process}.
\end{itemize}
\end{itemize}

\end{itemize}

Here we used the exponential distribution with $\bar{z}=0.2$ and an upper-length cutoff $g(z)= 1/\overline{z} e^{-z/\overline{z}} \Theta(z) \Theta(10 -z)$, where $\Theta$ is the Heaviside function.
We always used $\eta_u=1$.

Above $V^c$ we consider that the stationary regime is reached when the relative variation between the last three values of $\sigma$ does not exceed $1\%$. 
Below $V^c$ we ask that the stress drops $\Delta\sigma$ of three consecutive Global Shocks change less than $1\%$. 
Finally the critical value $V^c$ is found by decimation: starting with a very small $V_{min}=1e^{-4}$ and very large $V_{max}=1000$, simulations are repeated between the two boundaries until the relative difference between those two boundaries is less than $1\%$. This means that a larger absolute tolerance is allowed for larger values of $V^ c$.

\begin{figure}
 \label{figapp:convegence}
\includegraphics[width=0.48\textwidth]{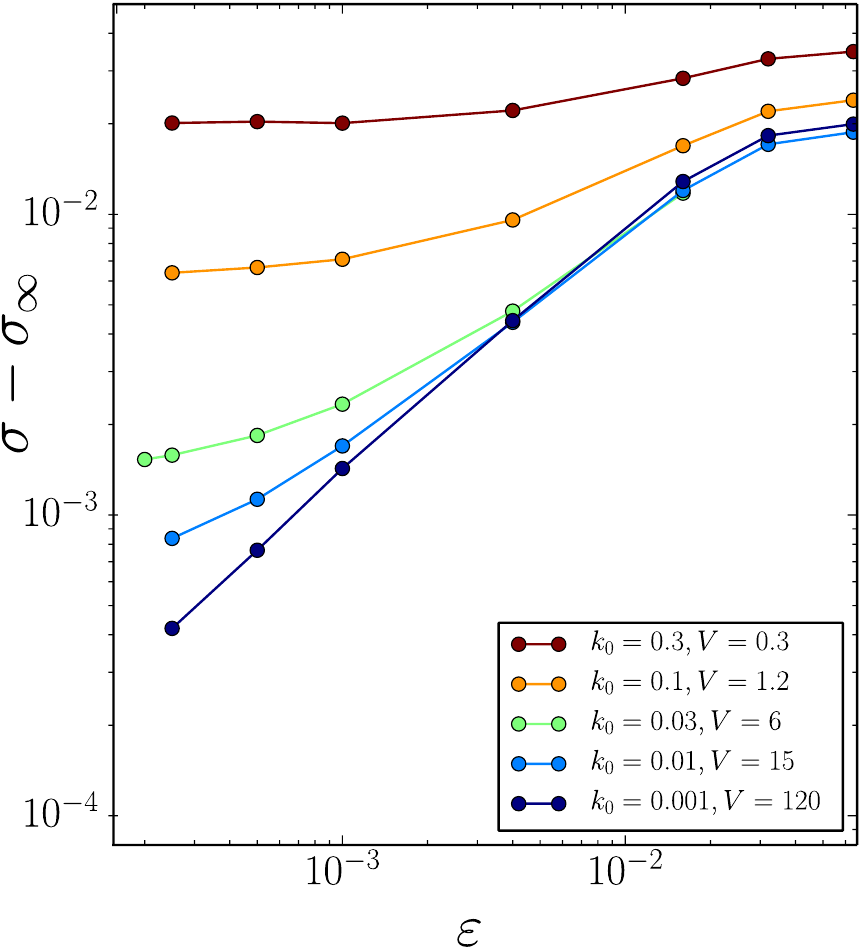}
  \caption{
  (Color online)
Convergence of the stress to its limit value $\sigma(V)$, as the binning $\varepsilon$ decreases. We selected velocities $V$ close to but larger than $V^c$ for the five values of $k_0$ studied (from top to bottom: decreasing $k_0$ and increasing $V$). Convergence is reached for the three largest values of $k_0$.
}
\end{figure}

An issue limiting higher precision is the presence of numerical instabilities when approaching to the critical point. 
In Fig.~\ref{figapp:convegence}
 the dependence of $\sigma$ on the binning $\varepsilon$ is shown. 
For large values of $k_0$ ($0.3, 0.1$ and $0.03$), it is clear that we converge towards a plateau $\sigma(V)$ when the binning decreases.
The small values of $k_0$ represent a challenge, because the precision required from the algorithm is $\sim \overline{z} k_0$. However, comparing the expected values (dashed lines) and the behavior at larger $k_0$'s, it can be expected that smaller binnings would produce the expected results.

\section{Analytical solution for the stationary regime} \label{stationary}

Let us now solve the model of Fig. \ref{A3_1}(b).
We call $f^0(t)$, $f^1(t)$, and $f^2(t)$ the forces coming from the branches with the $k_0$, $k_1$, and $k_2$ springs. The coordinate $h_i$ will jump
to $h_i+\overline{z}$ each time 
\begin{equation}
f^0(t)+f^1(t)+f^2(t)=f^{th}
\label{jump}
\end{equation}

Between jumps, the different forces behave as 
\begin{eqnarray}
f^0(t)&=&f^0_a+k_1Vt\\
f^1(t)&=&f^1_a+k_0Vt\\
f^2(t)&=&V\eta+(f^2_a-V\eta_u)\exp\left (-\frac{k_2t}{\eta_u}\right )
\end{eqnarray}
where the $a$ sub-indexes on the right indicate values of the forces right after the jump. This expression hold until the
next jump that occurs at time $t\equiv \overline{z}/V$. At this moment, the forces
must satisfy Eq. (\ref{jump}), and we obtain
\begin{eqnarray}
f^0_b&=&f^0_a+k_1\overline{z}\\
f^1_b&=&f^1_a+k_0  \overline{z}\\
f^2_b&=&V\eta+(f^2_a-V\eta_u)\exp\left (-\frac{k_2  \overline{z}}{V\eta_u}\right )
\end{eqnarray}
where the $b$ sub-index stands for the values right before the jump. In addition, we have $f^1_b=-f^1_a$ since the average value of $f^1$ must vanish. Also, since the dashpot is rigid at the jump, we get $f^2_b-f^2_a=\overline{z}k_2$. From all these equations all forces ($ f^0_b,f^0_a,f^1_b,f^1_a
f^2_b, f^2_a$) can be calculated. In particular we are interested in the average friction force $\sigma$ which is given by $\sigma=(f^0_a+f^0_b)/2$. Through a straightforward elimination procedure we get the Eq.(\ref{solucion}) given in the main text.

\end{document}